# Gradient polaritonic surface with space-variant switchable light-matter interactions in 2D moiré superlattices

Zhen-Bing Dai,[1,3]† Hua Fan,[2]† Vyacheslav Semenenko,[4] Xinyu Lv,[1] Lu Wen,[1] Zhen Zhang,[1] Shijie Fang,[2] Vasili Perebeinos,[4] Yue Zhao,[2]* Zhiqiang Li[1]*

**Abstract**
Polaritons in two-dimensional (2D) materials provide unique opportunities for controlling light at nanoscales. Tailoring these polaritons via gradient polaritonic surfaces with space-variant response can enable versatile light–matter interaction platforms with advanced functionalities. However, experimental progress has been hampered by the optical losses and poor light confinement of conventionally used artificial nanostructures. Here we demonstrate natural gradient polaritonic surfaces based on superlattices of solitons—localized structural deformations—in a prototypical moiré system, twisted bilayer graphene on boron nitride. We demonstrate on–off switching and continuous modulation of local polariton-soliton interactions, which results from dramatic modifications of topological and conventional soliton states through variation of local strain direction. Furthermore, we reveal the capability of these structures to spatially modify the near-field profile, phase and propagation direction of polaritons in record small footprints, enabling generation and electrical switching of directional polaritons. Our findings open up new avenues towards nanoscale manipulation of light–matter interactions and spatial polariton engineering through gradient moiré superlattices.

[1]College of Physics, Sichuan University; Chengdu, Sichuan 610064, China.
[2]Department of Physics and Shenzhen Institute for Quantum Science and Engineering, Southern University of Science and Technology of China; Shenzhen 518055, China.
[3]Department of Physics, Sichuan Normal University; Chengdu, Sichuan 610066, China.
[4]Department of Electrical Engineering, University at Buffalo; Buffalo, NY 14260, USA.

*Corresponding author. Email: zhiqiangli@scu.edu.cn (Z.L.); zhaoy@sustech.edu.cn (Y.Z.)

†These authors contributed equally to this work.

**Introduction**
In 2D heterostructures, a moiré pattern can be formed by stacking two van der Waals (vdW) materials with a twist angle or lattice mismatch. The resultant moiré superlattices can strongly modify the electronic band structure of the system, giving rise to a range of emergent quantum phenomena (1-5). The physical properties of moiré materials depend critically on the shape and periodicity of the superlattice (1-4). Gradient structures with spatially varying moiré superlattices thus offer the possibility of continuously tuning the moiré effects and resulting rich electronic properties in a single device, which may be potentially used as a continuously tuned quantum simulation platform to explore a broad spectrum of physics (2), especially emergent phenomena not present in homogeneous moiré systems (6). Such intriguing possibilities are now becoming achievable due to recent advances in controllable moiré patterning techniques (7). Therefore, explorations of these structures may unlock the potential of moiré-gradient engineering for realizing novel electronic and optical properties.

Polaritons—hybrid light–matter excitations—in vdW materials exhibit strong field confinement, long lifetimes and broad tunability, holding great promise for exploring strong light-matter interactions at the nanoscale (8-11). Gradient polaritonic surfaces— thin media devised to exhibit space-variant (aperiodic) optical response— have long been pursued as a powerful platform to control vdW polaritons for the realization of a myriad of advanced nanophotonic functionalities such as 2D gradient index optics and transformation optics, planar near-field lensing and wavefront engineering in atomically thin layers (10,12), with applications in imaging, on-chip photonic circuitry and quantum nanophotonics (11). However, their experimental implementation using conventional artificial structuring approaches suffers from severe optical losses and intrinsic limitations (by the structure size) in light confinement (10), which are particularly restrictive for controlling ultraconfined low-loss vdW polaritons. These limitations inherent in artificial structures highlight the importance of exploring alternative approaches, such as the development of natural gradient polaritonic surfaces, which could provide unique advantages for polariton manipulation.

Moiré superlattices offer an exciting platform for tailoring vdW polaritons (13), leading to many emergent photonic and optoelectronic phenomena. Structurally, moiré heterostructures feature atomic-scale lattice reconstruction at large periods (14-16), resulting in the formation of domains separated by superlattices of quasi-one-dimensional (1D) structural boundaries with concentrated strain known as solitons (14-20) (or domain walls). A prominent example is minimally twisted bilayer graphene (mTBG), in which electronic states confined at the solitons (which is an essential feature of solitons) produce spatial modulations of the local optical conductivity, leading to polariton-soliton scatterings (21-24). Such interactions render soliton superlattices a natural photonic crystal for graphene plasmons (21). So far, only periodic networks of shear solitons (21, 25–27) have been studied. What awaits discovery is the uncharted potential of gradient superlattices of solitons—inhomogeneous soliton superlattices with spatially varying superlattice distortions—for exploring and controlling vdW polaritons.

Here, we demonstrate a type of natural gradient polaritonic surfaces based on gradient superlattices of solitons (or gradient soliton superlattices for short) in 2D moiré materials, using mTBG on hexagonal boron nitride (hBN) as an example. In such reconstructed moiré systems, spatial variations of moiré geometry (namely, distortions of the soliton superlattice) lead to local variations of soliton angle $\varphi$—the angle between the local strain direction and the soliton normal (14) (Fig. 1A). We discover that the soliton angle provides a critical and hitherto neglected degree of freedom for nanoscale on-off switching and modulation of local light-matter interactions, adding a new physical mechanism for polariton manipulation in 2D materials. We show that both topological and conventional soliton states and, consequently, the local optical conductivity at solitons can be substantially modified by $\varphi$, leading to switchable polariton-soliton interactions though variation of $\varphi$ as well as electrical gating. Enabled by such unique local switchability, gradient soliton superlattices can be exploited to spatially modify the near-field profile (wavefront), propagation and phase of polaritons at subwavelength scales, which allows generating and switching directional polaritons with electrical tunability. Due to the ultrastrong polariton confinement and the narrow width of solitons, these versatile functionalities can be achieved within a record small footprint of sub-100 nm in our structure, up to 100 times smaller than previous artificial structures used for spatial control of polaritons (10, 28).

## Results
### Infrared (IR) nano-imaging of gradient soliton superlattices

mTBG hosts triangular superlattices of solitons with a width of 6 to 10 nm (14, 15) (Fig. 1B, C), which separate domains of energetically favorable AB/BA Bernal stacked regions with different in-plane displacement vector between the two graphene sheets $\boldsymbol{u}$. The soliton is associated with a local change in $\boldsymbol{u}$ (namely, strain $\boldsymbol{\Delta u}$) with saddle point stacking in the middle, with two limiting cases being tensile ($\varphi=0°$) and shear ($\varphi=90°$) solitons. We investigate the polaritonic response of mTBG/hBN heterostructures using nanoscale IR imaging (29, 30) at room temperature ($T$=298 K) (Fig. 1D). In these experiments, the metallized tip of an atomic force microscope (AFM) is illuminated by an IR laser with frequency $\omega$. Acting as an optical antenna, the tip concentrates the incident field to a nanoscale light spot (~20 nm) between its apex and the sample underneath, which can excite polaritons. The behaviors of polaritons depend crucially on the optical conductivity and hence the intraband, interband and many-body effects in the host or surrounding medium (8) thus providing a sensitive means for probing physical properties at the nanoscale. The local near fields are backscattered by the tip into the far field and their amplitude $s(\omega)$ and phase are measured simultaneously with topography. Recording the near-field signals while scanning the tip generates nanoscale near-field images. We studied devices consisting of exposed mTBG on hBN on graphite(gate)/SiO$_2$/Si substrate (Methods).

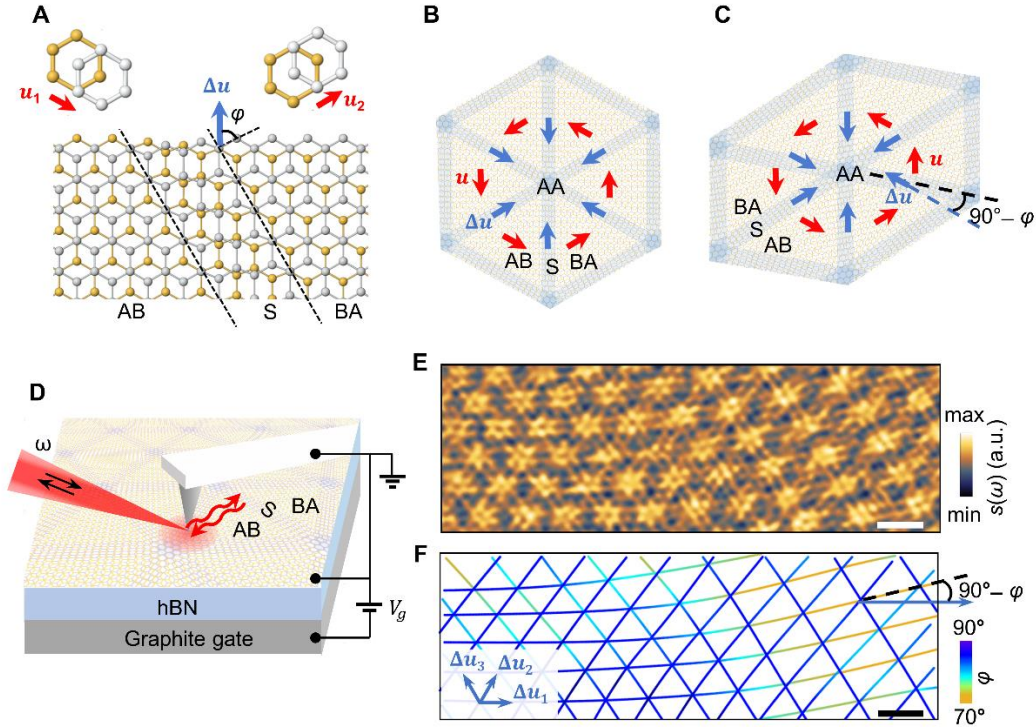

**Fig. 1. Infrared nano-imaging of gradient soliton superlattices and soliton angle variations in mTBG.** (**A**) Schematic of a soliton (S) in bilayer graphene separating AB and BA domains. Each domain is associated with a displacement vector between the two graphene sheets ($u_1$ and $u_2$), and the soliton is associated with a local change in $u$ (strain $\Delta u$). The soliton angle $\varphi$ is defined as the angle between the local strain direction and the soliton normal. (**B** and **C**) Illustration of mTBG with perfect sixfold symmetry (B) and with lattice distortion (C), with solitons shown as shaded regions. The vectors $u$ ($\Delta u$) associated with the six domains (solitons) intersecting at one AA region are successively rotated by 60°, as shown by red (blue) arrows. (**D**) Schematic of the IR nano-imaging experiment and device structure, mTBG/hBN/graphite(gate). Black arrows denote infrared light of frequency $\omega$ incident on and backscattered by an AFM tip. The red arrows represent the propagation direction of the polaritons. (**E** and **F**) Near-field amplitude $s(\omega)$ image of a typical sample (device A) acquired at $\omega$=1530 cm$^{-1}$ and a gate voltage of $V_g$=−6 V (E) and the corresponding local soliton angle map (F). Solitons shown as lines in (F) are visible as bright double-line features in (E). A finite angle

90°−$\varphi$ between $\Delta\boldsymbol{u}$ and the soliton is indicated in (C) and (F). a.u., arbitrary units. Scale bars, 200 nm.

Figure 1E shows the near-field amplitude $s(\omega)$ image of a representative sample (device A) acquired at $\omega$=1530 cm$^{-1}$ and a gate voltage of $V_g$=−6 V (carrier density ~6×10$^{12}$ cm$^{-2}$), in which the soliton network in mTBG manifests as a triangular network of double-line features centered at the solitons (24). Stronger contrast is observed at the intersections of solitons (AA stacking). No corresponding features can be seen in the AFM topography images acquired simultaneously with the near-field images. The spacing of the double lines decreases with increasing $\omega$ within the upper Reststrahlen band of hBN (~1360–1620 cm$^{-1}$). At such high doping, hyperbolic plasmon–phonon polaritons (31) are excited in the system, which originate from the hybridization of plasmons in mTBG with phonon polaritons in hBN (32, 33). Because these highly confined polaritons are very sensitive to the electrodynamics of the surrounding media, local optical conductivity changes across the solitons in mTBG (21, 22) can lead to effective scattering of these excitations (24). Specifically, the double-line features in the nano-IR image—corresponding to two peaks in the near-field amplitude line-profile $s(x)$ across the solitons ($x$ is the coordinate normal to the solitons) —are fringes produced by the interference between the tip-launched outward propagating polariton waves and those back-reflected by the solitons (Fig. 1D). The observation of only two peaks in $s(x)$ is attributed to polariton damping in these structures, as numerical simulations demonstrate that lower damping can yield additional peaks in the interference profile (24).

In reconstructed moiré superlattices, moiré gradients manifest as distorted soliton lattices characterized by spatial variations of soliton angle. The triangular soliton network in Fig. 1E is gradually distorted going toward the right region, which is caused by extrinsic heterostrain (differential interlayer strain) in mTBG. As shown by recent transmission electron microscopy experiments, all solitons in both perfect and distorted moiré superlattices (Fig. 1B, C) are associated with one of the three distinct $\Delta\boldsymbol{u}$ directions, $\Delta\boldsymbol{u}_1$, $\Delta\boldsymbol{u}_2$ and $\Delta\boldsymbol{u}_3$, which are successively rotated by 60° (14, 34). Such rotation is dictated by the successive 60° rotation of $\boldsymbol{u}$ in adjacent AB and BA domains. Shear solitons have the lowest energy per unit length compared with other types of solitons (14), so their formation is energetically favorable during the lattice reconstruction. For moiré pattern with sixfold rotational symmetry, all solitons are shear ones, which minimizes the total energy (Fig. 1B). Therefore, the nearly perfect sixfold moiré pattern observed in the left region of Fig. 1E allows us to estimate the three $\Delta\boldsymbol{u}$ directions as the three orientations of the solitons with an estimated accuracy of ±2°, from which the 2D soliton angle map $\varphi(\boldsymbol{r})$ (here $\boldsymbol{r}$ is the position within the soliton) can be obtained (Fig. 1F). Interestingly, we find that only one soliton array associated with $\Delta\boldsymbol{u}_1$ significantly deviates from shear strain in the right region, whereas the other two soliton arrays remain predominantly shear to minimize the total energy. As shown in Fig. 1C and 1F, the domain shape is closely related to the soliton angle of the non-shear solitons in distorted moiré patterns, in which lattice distortion induces a finite angle 90°−$\varphi$ between $\Delta\boldsymbol{u}$ and the soliton.

**Switching local polariton-soliton interactions with soliton angle**

Nano-imaging experiments reveal that soliton angle dramatically affects the local polariton-soliton interactions. Figure 2A shows the evolution of line-profiles $s(x)$ across single solitons at $x=0$ as a function of $\varphi$ at low doping, which is extracted from the nano-IR image of a typical sample (device B, which exhibits a broad range of $\varphi$) with strongly distorted superlattices acquired at $\omega=1540$ cm$^{-1}$ and $V_g=0$ (see Supplementary Materials section 2 and fig. S3). The $s(x)$ profiles are narrow due to the short polariton wavelength at this laser frequency, which allows one to study polariton-soliton scattering for single (isolated) solitons without significant influence from other solitons. At low doping, the polariton interference pattern in $s(x)$ for $\varphi\approx90°$ exhibits a dip feature centered at the soliton (fig. S2) corresponding to the dark feature in Fig. 2A. Remarkably, such feature gradually diminishes with decreasing $\varphi$ and vanishes below $\varphi\approx70°$. Since polaritons back-scattered by the solitons are required for producing the interference features, the flat profile in $s(x)$ for small $\varphi$—corresponding to no polariton interference—indicates that the polaritons pass through such solitons with negligible back-scattering (Fig. 2B). Therefore, the experimental data in Fig. 2A demonstrates on–off switching and continuous modulation of polariton-soliton interactions by $\varphi$ at low doping (Fig. 2D). On the other hand, polariton interference occurs at all solitons with $\varphi$ in the range of 65°–90° at $V_g=-6$ V as shown in Fig. 1E, F and fig. S4, enabling active on–off switching of polariton-soliton interactions by gating for $\varphi$ below ~70°. Similar on-off switching behavior with $\varphi$ has been observed in device A (fig. S4) as well as all devices with a sufficient range of $\varphi$ that we have studied.

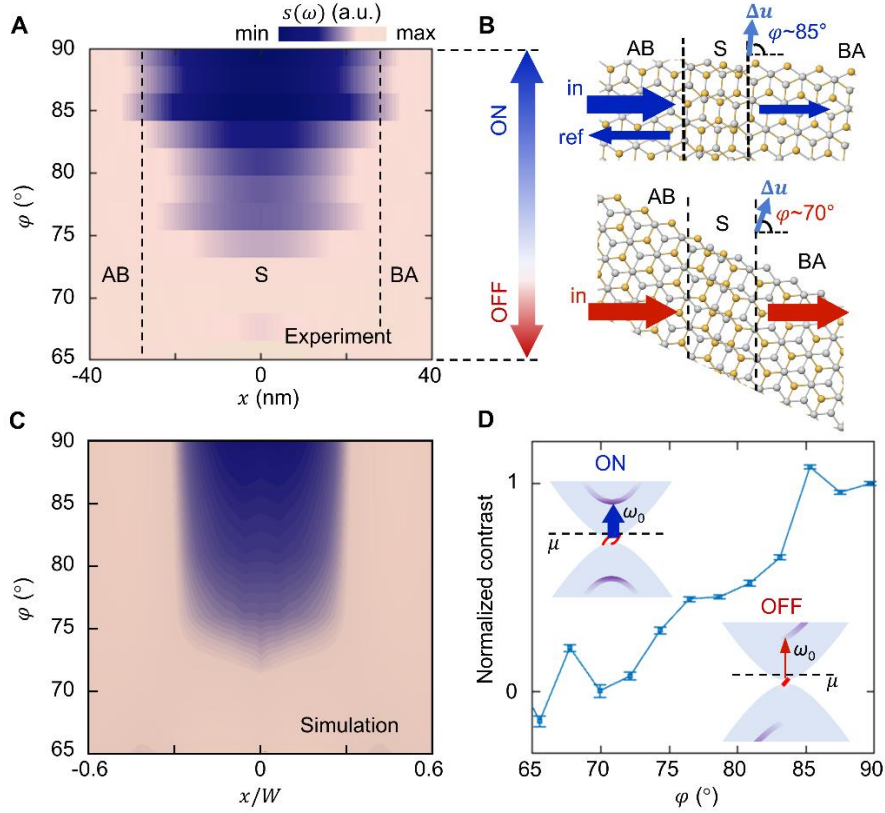

**Fig. 2. Switching on/off local polariton-soliton interactions with soliton angle.** (**A**) Experimental near-field amplitude $s(x)$ profiles across single solitons as a function of $\varphi$ acquired at $\omega$=1540 cm$^{-1}$ and $V_g$=0 V for a typical sample (device B). The labeled regions reflect the width of the interference feature. (**B**) Illustration of polariton (represented by arrows) scattering by solitons. In "ON" states, the incident polaritonic waves (indicated as "in") launched by the tip are partially reflected (indicated as "ref") by the soliton. In "OFF" states, the incident polaritons pass through the soliton with no interaction (reflection). (**C**) Simulated $s(x,\varphi)$ profiles at low doping based on continuum model calculations displayed in normalized coordinate $x/W$, where $W$ is a width beyond which the simulated profiles become flat. (**D**) Normalized contrast $\Delta s(\varphi)/\Delta s(90°)$ as a function of $\varphi$ extracted from (A), where $\Delta s$ is near-field signal contrast between the soliton and AB regions. The error bars represent standard errors. (Insets) Schematic depiction of electronic band structures of solitons for ON/OFF states. Shaded regions represent band continua of the bulk. Topological chiral 1D states are depicted in red. High-energy electronic states localized at the soliton are depicted by blue lines, with their darkness representing density of states. The evolution of optical transitions around $\omega_0$≈0.2 eV indicated by the arrows (with their width representing spectral weight) as a function of $\varphi$ is responsible for the observed on–off switching behavior. $\mu$, Fermi energy.

The soliton angle and gating dependence of the observed effects can be fully captured by theoretical modeling, which includes electronic structure calculations and numerical simulations of the near-field profiles $s(x)$ (Methods). Our calculations reveal strong $\varphi$-dependent band structure and local optical conductivity $\sigma(\omega, x) \equiv \sigma_{xx}(\omega, x)$ for solitons in bilayer graphene (figs. S5 and S6). The simulated $s(x,\varphi)$ profile at low doping (Fig. 2C) reproduces all experimental features and their $\varphi$ dependence in Fig. 2A. In particular, the flat $s(x)$ profiles around $\varphi \sim 70°$ are caused by the nearly flat $\sigma(x)$ profiles for such solitons, which make them transparent to the polaritons. At high doping, the simulated $s(x)$ exhibits a strong double-peak feature for all solitons with $\varphi > 65°$ (fig. S6), consistent with the experimental data (Fig. 1E).

The observed switchable polariton-soliton interactions arise from continuous evolution of soliton states with $\varphi$ (Fig. 3, figs. S5 and S6), which can be qualitatively understood as the superposition of local 2D band structures $E(k_x, k_y)$ at various locations across the soliton projected to the 1D momentum axis parallel to the soliton with a projection angle $\varphi$ (see Supplementary Materials section 3). All types of solitons host a pair of topological chiral 1D states (35, 36), which gradually move closer to one another until basically merging together and then move apart as $\varphi$ decreases from 90°. In addition, we find conventional bound soliton states outside the band continua and high-energy soliton states inside the band continua. The evolution of topological as well as conventional soliton states with $\varphi$ profoundly modifies the interband transitions involving these states and hence the local $\sigma(x)$. In particular, the observed $\varphi$ dependence of $s(x)$ at low doping (Fig. 2A) is caused by a resonance around 0.2 eV in the $\sigma(\omega)$ spectra for solitons shown in fig. S5 (see Supplementary Materials sections 3 and 4.4). The primary contribution to this resonance is from optical transitions from topologically non-trivial states to high-energy soliton states as indicated in Fig. 2D, which occur within a narrow energy window around the resonance. In addition, transitions from bulk states in the valence band to high-energy soliton states occur in a broader range of higher energies, contributing to the high-energy shoulder of the resonance peak (fig. S5). These assignments are further evident from the gradual suppression of the resonance peak as $\varphi$ decreases from 90° (fig. S5), resulting from the disappearance of high-energy soliton states in the band structure with decreasing $\varphi$ (Fig. 3). At high doping, the pronounced double-peak feature in $s(x)$ for all $\varphi$ is caused by the strong resonance peak in real part of $\sigma(\omega)$ around 0.2 eV in all cases (fig. S6). The simulated $s(x,\varphi)$ profiles exhibit similar behaviors for $\omega$ within the upper Reststrahlen band of hBN, in which our experiments were carried out. Besides this frequency range, our calculations show that similar switching of polariton-soliton interactions by $\varphi$ can be realized at high doping for $\omega \sim 650–900$ cm$^{-1}$ (see Supplementary Materials section 5 and fig. S7) due to the dramatic $\varphi$–dependence of a resonance around 0.1 eV in $\sigma(\omega)$, providing opportunities for manipulating a variety of vdW polaritons such as plasmon polaritons in mTBG and hybrid plasmon–phonon polaritons in mTBG/hBN and mTBG/α-MoO$_3$ (8–10).

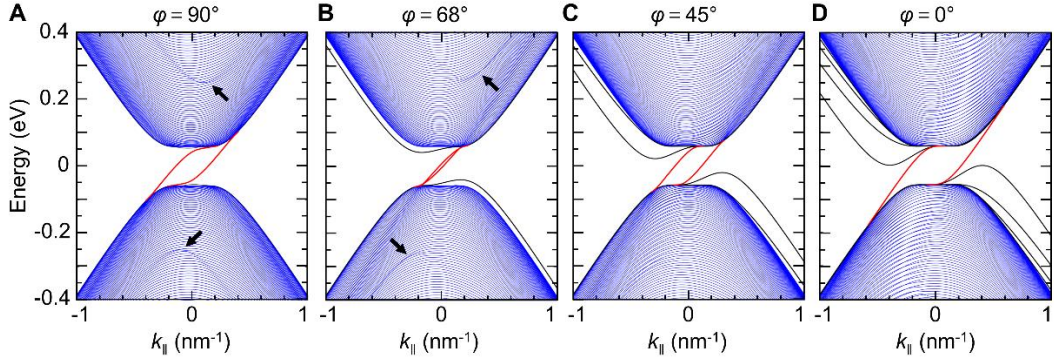

**Fig. 3. Electronic band structure of a single soliton in bilayer graphene at representative soliton angles.** (**A** to **D**) Electronic structure under a representative interlayer bias $V_i$=120 meV for the K valley. Inside the bandgap, the two topological chiral 1D states (depicted in red) gradually move closer to one another and then apart as the soliton evolves from shear to tensile (A to D). Conventional bound soliton states (black curves) exist outside the band continua of the bulk (depicted in blue) for non-shear solitons. Dispersing branches in darker color (indicated by black arrows) inside the band continua are high-energy states confined to the soliton. $k_\parallel$, momentum along the soliton.

**Tailoring polaritons through gradient soliton superlattices**

The unique tunability by soliton angle variations allows one to tailor the polaritonic fields at subwavelength scales exploiting the spatial profile of $\varphi(r)$ in gradient soliton superlattices. In particular, certain profiles can generate directional polaritons. Figure 4A and 4B displays the $\varphi(r)$ map and near-field image acquired at $\omega$=1500 cm$^{-1}$ and $V_g$=0 V for device B, respectively. The image provides a visualization of the interference between propagating polaritons triggered by the soliton network (21), since the polariton wavelength is much larger than the moiré period at this $\omega$, which leads to polaritonic interference due to interactions with multiple solitons and enables subwavelength tailoring of the polaritons. Based on Fig. 4A, the striped feature in the left dashed region in Fig. 4B reflects directional polariton propagation predominantly induced by interactions with the vertical soliton array with $\varphi$≈90°, because the other two soliton arrays with small $\varphi$ are in the "OFF" states (Fig. 2A). To corroborate this, we simulate the $s(\omega)$ images using a superposition model (21) considering polariton interference effects induced by one array of nearly shear solitons (see Supplementary Materials section 4 and fig. S3E), which shows excellent agreement with the observed striped feature. The simulated field distributions of polaritons (Methods) in this region at low doping (Fig. 4D) exhibit a highly directional wavefront perpendicular to the solitons, with the electric field mainly concentrated on the solitons due to strong collective near-field coupling.

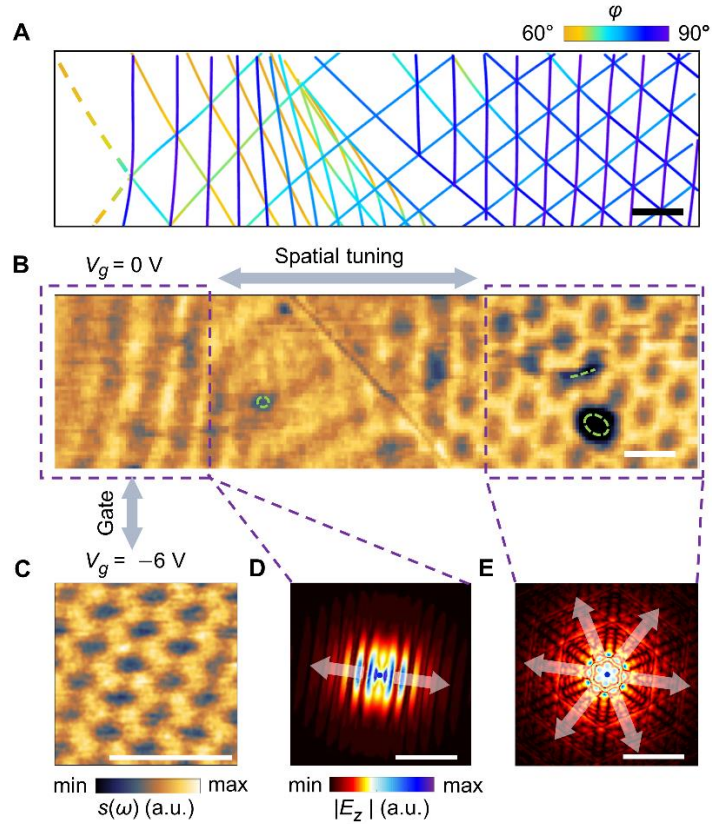

**Fig. 4. Tailoring polaritons through spatial variations of soliton angle in gradient soliton superlattices.** (**A** and **B**) Soliton angle map (A) and nano-IR image acquired at $\omega$=1500 cm$^{-1}$ and $V_g$=0 (B) for device B. Locations of defects determined from AFM topography are indicated by green dashed circles and lines in (B). (**C**) Nano-IR image acquired at $\omega$=1530 cm$^{-1}$ and $V_g$=−6 V for device C with strong lattice distortion similar to the left dashed region in (B). (**D** and **E**) Simulated field distributions (absolute value of the z-component of the electric field |$E_z$|) of polaritons for the left (D) and right (E) dashed regions in (B), as launched by a dipole (indicated by the blue dot) placed 100 nm above the sample surface. (D) and (E) illustrate polariton propagation under the influence of one and three array(s) of nearly shear solitons, respectively, with the arrows indicating the predominant propagation directions of polaritons. Scale bars, 200 nm.

The near-field profile of such directional polaritons can be actively switched into hexagonal shapes by electrostatic gating. The gate-dependent nano-IR image of a similar sample with strong lattice distortion (device C) evolves from striped features at $V_g$=0 (fig. S8A) to a hexagonal pattern of dark spots at AA regions with increasing gate voltage (Fig. 4C). As corroborated by simulations (fig. S8C), such hexagonal pattern arises from interference of propagating polaritons triggered by interactions with all three soliton arrays, because the "OFF" states for soliton arrays with small $\varphi$ are switched on by gating

(fig. S6). Correspondingly, the polaritons exhibit a distorted hexagonal wavefront complying with the symmetry of the soliton lattice (fig. S8D).

Furthermore, we demonstrate that the soliton angle profile $\varphi(r)$ in gradient soliton superlattices enables spatial modification of the wavefront, propagation direction and phase of polaritons within a small footprint of sub-100 nm. In a typical sample (device B) where the soliton lattice changes gradually from highly distorted to nearly sixfold pattern (Fig. 4A), the nano-IR image (Fig. 4B) evolves from a striped pattern (left region) to a hexagonal pattern (right region), which is induced by interactions with one and three soliton arrays (fig. S3, E and F), respectively. This image illustrates a spatial transition from highly directional (Fig. 4D) to hexagonal wavefront (Fig. 4E) induced by the spatial dependent polariton-soliton interactions governed by $\varphi(r)$ in Fig. 4A. All devices with soliton angle distributions similar to device B (for example, those shown in figs. S8, S9B and S12) exhibit similar behaviors. In addition, our data show that directional polaritons can travel parallel, perpendicular or oblique to a boundary (see Supplementary Materials section 5 and fig. S9) between highly distorted and nearly sixfold soliton lattice, depending on the orientation of the shear soliton array that governs the directional wavefront. Such behaviors provide possibilities for polariton steering. Our simulations also show that the phase of the plasmon polaritons in mTBG can be continuously modified by varying the soliton angle (see Supplementary Materials section 5 and fig. S10), pointing to the potential of using $\varphi(r)$ for spatial phase modulation. These functionalities enabled by gradient soliton superlattices can be realized in a broad IR frequency range (figs. S5 and S6), offering opportunities for manipulating a plethora of vdW polaritons.

**Discussion**

We have demonstrated that moiré gradients (soliton angle profiles) intrinsically govern the light-matter interactions in reconstructed gradient moiré superlattices, enabling locally switchable polaritonic responses and versatile functionalities. The functionalities demonstrated here are unachievable in previous artificial gradient structures used for spatial control of polaritons (10, 28), which suffer from severe optical losses (37) and very limited light confinement (confinement factor $\beta\sim3$ and hence large footprint) (28). In contrast, gradient moiré superlattices studied here can host hybrid plasmon–phonon polaritons with record long lifetimes (a few picoseconds) while maintaining their ultrastrong light confinement ($\beta>100$, enabling ultracompact footprint) (8-10), which overcomes the drawbacks of artificial structures and provides substantial advantages for planar nanotechnologies. Moreover, the capability of gradient moiré structures is not limited to the specific soliton distributions studied in our work, but goes beyond. Other different spatial distributions can enable diverse functionalities based on the polaritonic effects demonstrated here.

Our findings unlock the enormous potential and intrinsic capability of 2D moiré superlattices to serve as gradient polaritonic surfaces. Although the soliton distributions are not controlled in our samples, controllable creation and fine tuning of gradient soliton networks (including and beyond those in our samples) have become achievable due to continued recent advances in fabrication techniques, such as moiré patterning with in-plane bending of monolayer ribbons [forming gradient soliton networks in regions with twist angle <0.5° (34)] (7, 38), piezoelectric-based strain engineering (18) and AFM manipulation of twistable TBG devices (19, 39). These timely advances underscore the broad implications of our findings from both fundamental and applied perspectives (40, 41). Combined with such controllable gradient soliton superlattices, the intrinsic switching and tuning capabilities demonstrated in our study open up new avenues to tailor local light-matter interactions for nanoscale spatial polariton engineering with unprecedented functionalities. This mechanism for polariton control can be extended to soliton networks in diverse 2D moiré systems for exploring and exploiting a rich variety of polaritons (8-11) due to the inherent electric polarization of solitons (17) facilitating their interactions with polaritons. Such new class of gradient polaritonic surfaces exhibits combined merits of low loss, strong light confinement, ultracompact footprint and advanced functionalities, offering significant advantages over artificial nanostructures. Fundamentally, the strong tunability of soliton states and soliton-polariton interactions demonstrated here is expected to profoundly impact a wealth of photonic excitations and effects in moiré soliton superlattices, such as topological chiral plasmons (25), interactions between polaritons and correlated soliton states (20), and polaritonic probe of emergent effects (8, 13), thus providing new opportunities for exploring moiré photonics.

## Materials and Methods

### Device fabrication and characterization

The mTBG/hBN devices used in this study were prepared by a dry transfer method (42). The original graphene piece was cut into two parts by AFM probe before the transfer. Using stamps made from poly(bisphenol A carbonate) (PC)/polydimethylsiloxane (PDMS) at a temperature of approximately 100 °C, we picked up the left and right parts of the monolayer graphene (43), with a setpoint twist angle of ~0.1°, along with hBN (15–40 nm thick) and few layer graphene (FLG, or, graphite) crystals to form mTBG/hBN/FLG heterostructures. Lattice relaxation leads to soliton superlattices separating AB/BA Bernal stacked domains in mTBG. The assembly was released onto a $SiO_2$/Si substrate at a temperature of 180 °C, followed by removal of the PC stamp through a solvent wash using chloroform and iso propyl alcohol (IPA). Electrical connections to the mTBG and FLG gate were established through subsequent e-beam lithography and metal evaporation. Finally, the device was cleaned using contact mode AFM (~0.2 μN loading force) to ensure a residue-free surface. For some devices, we used a thin mica substrate with graphite and hBN stacks for further pickup of twisted bilayer graphene, so that the final device has an effective backgate as well as a fresh clean surface (see Supplementary Materials section 1 and fig. S1 for details). In the main text, we show data from devices with 19 nm-thick hBN serving as the gate dielectric, and the carrier density as function of gate voltage $V_g$ is 1.05±0.06×10$^{12}$ cm$^{-2}$/V.

As exposed surfaces are favored for SNOM, the devices we studied are without top hBN dielectrics. To evaluate the initial doping levels of our devices after the complicated fabrication flow, we performed scanning tunneling microscopy (STM) on a few devices. Our measurements have confirmed high quality surface with extremely low level of intrinsic doping on the sample surfaces. fig. S1G is the topographic image with the atomic resolution taken on a monolayer region of device STMS1, indicating an ultra-clean surface. The scanning tunneling spectroscopy curves at various locations ($V_g$=0 V) all exhibit a V shape with a nearly zero Dirac point, as shown in fig. S1H, within the range from −1 to +4 meV.

**Infrared nano-imaging measurements**

All infrared nano-imaging measurements were performed using scattering-type scanning near-field optical microscopy (SNOM) (29, 30), which is based on an AFM operating in tapping mode using cantilevered metallic tips with an apex radius of ~25 nm and tapping frequencies of ~270 kHz. An infrared laser beam is focused onto the apex of the tip. The enhanced optical field at the tip apex interacts with the sample underneath and modifies the local near field. The resulting near field is backscattered by the tip into the far field, which is recorded with a pseudo-heterodyne interferometric detection module to extract both the scattering amplitude $s(\omega)$ and phase $\psi(\omega)$ of the near-field signal. To suppress background contributions to the backscattered near-field signal, the detected signal is demodulated at the third harmonic of the tip tapping frequency.

**Band structure and local optical conductivity for solitons in bilayer graphene**

Our calculations for a single infinitely long soliton with soliton angle $\varphi$ separating bilayer graphene with AB and BA stacking are based on a continuum model with 4 × 4 Dirac-type Hamiltonian $H(k_x, k_y)$ for bilayer graphene with a spatial dependent interlayer displacement vector $\boldsymbol{u}(x)$ (22, 44, 45), where $x$ is the direction normal to the soliton. The momentum $k_y$ remains a good quantum number. The momentum $k_x$ perpendicular to the soliton is replaced by the operator $-i\partial/\partial x$. The resulting real space Hamiltonian $H(x, k_y)$ is solved numerically on a 1D grid along $x$ direction to obtain electronic band structures as shown in figs. S5 and S6. In the cases of shear and tensile solitons, we obtain similar results as reported in previous studies (22). Next, the eigenvalues ($E_m$, $E_n$) and eigenfunctions ($|u_m\rangle$, $|u_n\rangle$) of the Hamiltonian are used to compute the nonlocal optical conductivity $\Sigma(x, x')$ employing the Kubo formula (22). Then the local optical conductivity can be obtained from $\sigma(x) \equiv \int \Sigma(x, x')\, dx'$. The parameters used in the formula are chemical potential $\mu$, temperature $T$, frequency $\omega$, interlayer bias $V_i$ and damping rate $\eta$. The details of these calculations and the dependence of the results on parameters (46) are discussed in Supplementary Materials section 3.

**Numerical simulations**

For a quantitative analysis of the observed soliton-angle- and gate-dependent near-field contrast, numerical simulations of the near-field line-profiles $s(x)$ across a single soliton are performed using two methods: a finite-element method (47, 48) and an analytical model (49), which produce consistent results. The structure in the simulations consists of bilayer graphene with a single soliton, hBN and SiO$_2$. In the finite-element method, the oscillating tip is approximated as an elongated conducting spheroid. A constant electric field in the $z$ direction, $E_0 \hat{\mathbf{z}}$, is used as a background excitation to induce the tip-sample near-field interaction, which further modifies the charge distribution along the spheroid surface and thus the dipole moment. Due to the elongated tip shape as well as the background field direction, only the dominant dipole moment in the $z$ direction, $p_z$, is considered. To simulate a spatial scan, the lateral position of the tip is varied. As for the second analytical method, near-field response is calculated from a dipole moment $\vec{p}$ of the tip averaged over its position oscillating along the direction normal to the sample. In both models, the oscillating dipole is demodulated to higher harmonics of the tip oscillation frequency to obtain the near-field amplitude and phase. The details of these simulations are discussed in Supplementary Materials section 4.

The polariton field distributions displayed in Fig. 4D are simulated by the finite-element method: we place a dipole (polarized along the $z$ direction) 100 nm above the sample surface to launch polaritons, and monitor the absolute value of the $z$-component of the electric field $|E_z|$ at the sample surface. The mTBG layer for simulating Fig. 4D contains one array of shear solitons. We perform calculations for bilayer graphene with equally spaced solitons with periodic boundary condition to obtain the $\sigma(x)$ profiles using similar parameters shown in fig. S5 (low doping). When the spacing between solitons is large enough, $\sigma(x)$ for each soliton is similar to the $\sigma(x)$ profile shown in fig. S5. The field distributions in Fig. 4E are simulated using a superposition model (21), using the field distributions in Fig. 4D as inputs for each of the three soliton arrays. The field distributions in fig. S8D are obtained using similar methods as Fig. 4E, except that the $\sigma(x)$ profiles for each soliton array are calculated using similar parameters shown in fig. S6 (high doping).